\documentclass[12pt]{iopart}
\usepackage{iopams}  

\begin{document}
\title[Classical Electrodynamics and Absolute Simultaneity]{Classical Electrodynamics and Absolute Simultaneity}
\author{Benjamin Nasmith}
\address{2 Canadian Forces Flight Training School, P.O. Box 5000 Stn Main, Moose Jaw, SK, Canada S6H 7Z8}
\ead{Ben.Nasmith@gmail.com}
\begin{abstract}
Maxwell's equations and the Lorentz force density are expressed using an alternative simultaneity gauge.  As a result, they describe electrodynamics for an observer travelling with a constant velocity through an isotropic medium.  If desired, one may regard the isotropic medium as an absolute space.  An observer at rest in the medium may be referred to as a preferred observer.  The medium appears electrically polarized to the moving observer.  Such an observer regards the speed of light as anisotropic by convention while sharing absolute simultaneity relations with the preferred observer.
\end{abstract}

%\submitto{\EJP}

\section{Introduction}
There seems to be a popular view among physicists and educators that one ought not to take absolute simultaneity too seriously in light of our current understanding of relativity physics.  
For example, the authors of \cite{Sche2002} study methods to help physics students overcome their intrinsic belief in absolute simultaneity. Why so?  
Perhaps the simplest reason is that the concept of simultaneity employed within the special theory of relativity (STR) is observer dependent.  
Different observers travelling with different velocities will disagree about whether two events occur simultaneously.  
This phenomenon, named the relativity of simultaneity, is often thought to count against the intuitive notion of a world-wide present. %that exists independent of one's individual motion.  
Upon closer inspection, however, it is difficult to see exactly why this is so. % how the concept of simultaneity employed within STR counts against the existence of a world-wide now.  
Consider the following two claims. First, the relativity of simultaneity within STR is never directly observed and is therefore experimentally indistinguishable from other unobserved simultaneity relations. 
Second, there are in principle at least two physically warranted candidates for the concept of absolute simultaneity.  
If these two claims are true, then one ought to be able to formulate relativistic mechanics and electrodynamics in terms of absolute, rather than relative, simultaneity. 

%%%%% 
Certain authors have already worked on these problems.  For example, both Anderson \etal \cite{Ande1998} and Sonego and Pin  \cite{Sone2009} provide detailed discussions of relativistic mechanics using alternative simultaneity relations. 
On the electrodynamics front, Tangherlini \cite{Tang2009}, Giannoni \cite{Gian1978}, and Rizzi \etal \cite{Rizz2004} discuss Maxwell's equations in tensor form using non-orthogonal coordinates.
%
%\footnote{The authors of \cite{Rizz2004} also discuss Maxwell's equations, although their final result has an ``unpleasant asymmetric form." These authors maintain that ``Maxwell's equations are covariant under synchronization changes, but optical anisotropy breaks down their standard symmetric form." } 
%
These coordinates permit absolute simultaneity relations and yield anisotropic wave equations governing the electromagnetic fields.  Commenting on Giannoni's work Anderson \etal write, 
\begin{quote}
These equations \ldots help to remove doubt as to whether the detailed dynamics of light propagation imposes a preference on any synchronization scheme. A student who has mastered the consequences of an anisotropic synchronization to this point has passed a significant threshold in understanding which will fortify against the recent erroneous trends in the literature \cite{Ande1998}.
\end{quote}
However, a similar derivation that does not require tensor calculus appears to be absent in the literature.  
Furthermore, Rizzi \etal suggest that the anisotropic Maxwell's equations suffer from an ``unpleasant asymmetric form" \cite{Rizz2004}.
In what follows, I will re-derive Maxwell's equations using merely an anisotropic wave equation and the conservation of electric charge.  This derivation reveals that an observer who considers themselves to be in absolute motion must account for the apparent electric polarization of the medium through which they travel.  This element is easily overlooked in the tensor formulation due to the distinction between the covariant and contravariant electromagnetic fields.  
It has also been overlooked in \cite{Pucc2005} where the author assumes, ``the constitutive relations in vacuum [are] valid not only in the privileged frame \ldots  but also in the moving frame."
Hopefully awareness of this effect will provide some clarity for those who wish to consider the compatibility of electrodynamics and absolute simultaneity.  

Before proceeding with the derivation in Section \ref{AnisotropicWE}, let's discuss the two claims made above.  
First, the relativity of simultaneity within STR is never directly observed and is therefore experimentally indistinguishable from other unobserved simultaneity relations. 
In order to illustrate this point, consider two observers who are momentarily collocated yet in relative motion as they pass each other; call them Alice and Bob.  
In introductory treatments of STR, one normally finds a discussion of how Alice and Bob observe the three phenomena of length contraction, clock retardation, and the relativity of simultaneity.  Unfortunately, the term `observe' is easily confused with what Alice and Bob would actually perceive visually.  Strictly speaking, Alice and Bob perceive the light arriving from various directions emitted by various distant events in the past.  As such, they cannot directly perceive distant events, such as a supernova.
Rather they perceive the incoming light from the supernova.  Terrell has shown that neither Alice nor Bob are able to `see' the Lorentz contraction as a result \cite{Terr1959}.
Furthermore, the only time Alice or Bob will actually see a clock tick as predicted by clock retardation is when it passes directly abeam them on its trajectory, due to the transverse Doppler effect.  
Rather than seeing the Lorentz contraction or clock retardation, Alice and Bob visually perceive the frequency, or colour, of incoming light and the direction from which it arrives \cite{Blat1988}.

%%%%%
How then do Bob's perceptions differ from Alice's given their relative motion?  
Bob's perceived light signals undergo Doppler shift and relativistic aberration compared to Alice's and vice-versa \cite{Blat1988}.  These two directly perceivable effects are quite intuitive since they also apply to sound signals.  
Blatter and Greber write, ``Doppler shift is easily witnessed near any street or railway line with fast moving vehicles and aberration with fast moving airplanes, where the sound seems to come from a direction somewhere behind the visible plane"\cite{Blat1988}. It is important to note that both Doppler shift and aberration are independent of clock synchronization procedures \cite{Roth1995}. 
Indeed, visual perception---a purely local observation---does not require multiple clocks and is therefore synchronization independent.  
As such, Alice and Bob cannot see the relativity of simultaneity either. 

What role then does simultaneity play?  Anderson and Stedman propose that the freedom to arbitrarily synchronize a system of distant clocks without influencing local phenomena, such as visual perception, may be understood as a gauge freedom \cite{Ande1992}. This view is further examined by Minguzzi who writes, 
\begin{quote}
Both [simultaneity and electrodynamics] have the mathematical structure of a gauge theory over a one-dimensional group \ldots Other analogies, like that between the Sagnac effect and the Aharonov-Bohm effect, or like that between magnetic forces and the Coriolis forces, become self evident in light of the gauge interpretation  \cite{Ming2003}.
\end{quote}
On this view, one cannot directly test for the relativity of simultaneity any more than one can ``test for the absolute zero of voltage in seawater" since simultaneity---like electric potential---is gauge dependent \cite{Ande1998}.  

%%%%%
In is important to realize that the speed of light depends on one's choice of simultaneity gauge.  Following Anderson and Stedman \cite{Ande1992}, let's consider a coordinate system $\left(\textbf{x}_0,t_0\right )$ in which the speed of light is isotropic with magnitude $c$.  Given the gauge freedom to adjust simultaneity arbitrarily, one may define a second coordinate system $\left(\textbf{x}_{\bkappa},t_{\bkappa}\right )$ as follows:
\begin{eqnarray}
\label{Ktrans}
\textbf{x}_{\bkappa}=\textbf{x}_{0}; \qquad t_{\bkappa}=t_{0}-\frac{\bkappa\cdot\textbf{x}_{0}}{c}.
\end{eqnarray}
In what follows $\bkappa$ will be a constant vector, sacrificing generality for simplicity.
A subscript $\bkappa$ will indicate that the quantity subscripted must be evaluated using the $\bkappa$ simultaneity gauge.  
A subscript $0$ will indicate that the quantity is to be evaluated using $\bkappa=0$ simultaneity.
\footnote{The vector magnitude $\kappa$ may be related to Reichenbach's $\epsilon$ parameter according to $\kappa=2\epsilon-1$ \cite{Ande1998}.}
Anderson and Stedman show that under this transformation the speed of light in the direction $\hat{\textbf{n}}$ is given by \cite{Ande1992}:
\begin{eqnarray}
\label{COneway}
\textbf{c}_{\bkappa}=\frac{c\hat{\textbf{n}}}{1-\bkappa\cdot\hat{\textbf{n}}}.
\end{eqnarray}
On this definition, the round-trip time for light to travel along a closed path remains independent of synchrony vector $\bkappa$: the average speed of light is $c$ \cite{Ande1992}. Of course, one cannot measure the one-way speed of light \cite{Spav2012} and therefore no experiment can detect the value of $\bkappa$, although claims to the contrary do arise occasionally \cite{Grea2009}.
 The special case where $\bkappa=0$ and light speed is isotropic represents Einstein's famous ``\textit{stipulation} which I can make of my own freewill in order to arrive at a definition of simultaneity [italics in the original]"\cite{Eins1961}.  Given its near universal use among physicists, I will refer to this special case as standard simultaneity in what follows.  To summarize the first claim, the relativity of simultaneity cannot be directly observed since no experiment can properly claim to observe the value of $\bkappa$ or the one-way speed of light.  The gauge nature of simultaneity permits alternative definitions.  For a very detailed exposition of this first claim, please refer to \cite{Ande1998}.

%%%
Let's consider the second claim given above: there are at least two physical candidates for the concept of absolute simultaneity.  The first option is to regard motion with respect to the cosmic background radiation (CMB) as absolute.  In this case, if Alice observes the CMB to be completely isotropic then she may regard herself as at rest and her $\bkappa=0$ frame of reference as preferred.  Bob will then find the CMB to be anisotropic due to aberration and Doppler shift regardless of what value of $\bkappa$ he chooses. Furthermore, Greber and Blatter write, ``the interaction of matter with the cosmic background radiation tends to bring the matter to rest in a preferred frame in which the radiation is isotropic"\cite{Greb1990}. As such, Bob will experience radiation pressure from the CMB tending to bring him to rest relative to Alice.  This provides some physical warrant to the view that Alice is a preferred observer.

%%%
The second option is to appeal to the instantaneous quantum correlations provided by the EPR experiment and Bell's inequalities. 
Craig and Smith write, ``The experimental confirmation of Bell's theorem by Aspect and others implies that there is an instantaneous, non-local, space-like relation of simultaneity that coincides with EPR causal correlations"\cite{Crai2008}.  This relation of simultaneity defines a preferred reference frame in which standard $\bkappa=0$ simultaneity overlaps with instantaneous quantum causal correlations \cite{Crai2001}. If Alice is at rest in such a preferred frame of reference then there is physical warrant to regard her as at absolute rest.  The fact that Alice cannot detect the preferred reference frame does nothing to prevent it from existing.  Rather, its existence rests on the experimental confirmation of Bell's theorem.  For additional discussion of these and other candidates for absolute simultaneity, please refer to \cite{Crai2008,Crai2001}.

In summary, I have briefly argued that the relativity of simultaneity is invisible to the observer and may be understood merely as a popular choice of simultaneity gauge, namely $\bkappa=0$ for all observers.  I have also provided two physically warranted candidates for the concept of absolute simultaneity.  In what follows I will suppose that Alice is at rest with respect to absolute space, however one chooses to physically define it, and that Bob is therefore in absolute motion.  Alice will use the $\bkappa=0$ simultaneity gauge and Bob will choose a non-zero value of $\bkappa$ such that he and Alice agree on which events are simultaneous.

Which value of $\bkappa$ should Bob use in order to share absolute simultaneity relations with Alice?  First, we know that if both Alice and Bob set $\bkappa=0$, then they can relate their observations using Lorentz transformations.  Second, the Lorentz transformations are known to consist of three steps: a Galilean transformation, a rescaling of the units of length and time, and a resynchronization of clocks \cite{Tang2009,Iyer2010}.  The first and third steps form groups although the second step does not \cite{Tang2009,Iyer2010}.  In order to ensure that Bob shares absolute simultaneity relations with Alice, he must simply invert the last step of the Lorentz transformation.  Therefore, if Bob is travelling with velocity $\textbf{v}$ relative to Alice, then he must set $\bkappa=-\textbf{v}/c$ in order to undo the last step of the Lorentz transformation.  By so doing, Alice and Bob will agree on which events are simultaneous and Bob will regard the speed of light to obey the following relation \cite{Ande1998,Sone2009,Tang2009,Gian1978,Ande1992,Spav2012,Iyer2010}:
\begin{equation}
\textbf{c}_{-v/c}=\frac{c\hat{\textbf{n}}}{1+\displaystyle{\frac{\textbf{v}}{c}}\cdot\hat{\textbf{n}}}.
\end{equation}

\section{Anisotropic Wave Equation}
\label{AnisotropicWE}
The task of the remainder of the paper is to work out electrodynamics from Bob's perspective where $\bkappa\ne 0$ and the speed of light is regarded as anisotropic.
In \cite{Hera2007}, Heras claims to derive Maxwell's equations using charge continuity alone.  Upon closer inspection, however, Heras implicitly requires that these charges are the sources of fields governed by the wave equation.  Charge continuity per se is not unique to electric charge.  Any field theory that enjoys local gauge invariance of its Lagrangian density will correlate to conserved Noether currents (see for example \cite{AlKu1991}).  Assuming mere charge continuity does not distinguish electrodynamics from other theories such as the Proca generalization or the Yang-Mills theory. To obtain Maxwell's equations one must assume \textit{electric} charge continuity.  

Essentially, Heras has shown that one may construct a model based solely on charge continuity and the wave equation.  This model may be tested experimentally to validate its two assumptions and fix the unknown constants.  I will follow a similar method to Heras in order to develop electrodynamics from Bob's perspective as an observer in absolute motion. In this section, I will develop an anisotropic wave equation with a particular solution.  In Section \ref{MaxwellEqs}, I will apply charge continuity to this anisotropic wave equation and recover Maxwell's equations.  In Section \ref{Dynamics}, I will derive the Lorentz force density and examine how one might fit the proposed model to experimental data.

Let's begin with the standard simultaneity wave equation:\begin{equation}
\label{WaveEqn}
\left (\frac{1}{c^2}\frac{\partial^2}{\partial t_0^2}-\nabla_0^2\right )\Psi=\psi.
\end{equation}
The homogeneous solutions to this equation travel with an isotropic velocity $c$.  This equation also has both retarded and advanced particular solutions.  In order to obtain an anisotropic generalization of Eq. \eref{WaveEqn}, one may make the following substitution based on Eq. \eref{Ktrans}:
\begin{equation}
\label{Koperators}
\nabla_0=\nabla_{\bkappa}-\frac{\bkappa}{c}\frac{\partial}{\partial t_{\bkappa}}; \qquad \frac{\partial}{\partial t_0}=\frac{\partial}{\partial t_{\bkappa}}.
\end{equation}
Therefore, the Laplacian may be written as follows:
\begin{equation}
\label{Klaplacian}
\nabla_0^2=\nabla_{\bkappa}^2+\mathcal{P}_{\bkappa}\frac{\partial}{\partial t_{\bkappa}}.
\end{equation}
Here I have introduced the following shorthand: 
\begin{equation}
\label{Pkdef}
\mathcal{P}_{\bkappa}= \frac{\bkappa^2}{c^2}\frac{\partial}{\partial t_{\bkappa}}-\frac{2}{c}\bkappa\cdot\nabla_{\bkappa}.
\end{equation}
Using Eqs. \eref{Koperators} and \eref{Klaplacian}, one may write Eq. \eref{WaveEqn} in anisotropic form:
\begin{equation}
\label{WaveEqnAni}\left (\frac{1}{c^2}\frac{\partial^2}{\partial t_{\bkappa}^2}-\nabla_{\bkappa}^2-\mathcal{P}_{\bkappa}\frac{\partial}{\partial t_{\bkappa}}\right )\Psi=\psi.
\end{equation}
The homogeneous solutions of this wave equation travel with the anisotropic velocity given by Eq. \eref{COneway}.

Let's consider the particular solution of Eq. \eref{WaveEqnAni}.  Only the retarded solution is discussed below since the advanced solution may be obtained in a very similar manner.  The retarded particular solution to Eq. \eref{WaveEqn} involves the following Green's function (for the derivation, please refer to \cite{Jack1999}):
\begin{equation}
\label{Green0}
 G(\textbf{x}'_0,t'_{0},\textbf{x}_0,t_0)=\frac{\delta\left (t'_0-t_0+\displaystyle{\frac{R_0}{c}}\right )}{4\pi R_0}.
\end{equation}
This expression uses the following notation: $\textbf{R}_0=\textbf{x}_0-\textbf{x}'_0$, $R_0=\left |\textbf{x}_0-\textbf{x}'_0\right |$, and $\hat{\textbf{R}}_0=\textbf{R}_0/R_0$.  
To obtain the Green's function for the anisotropic wave equation Eq. \eref{WaveEqnAni} one simply needs to apply the transformation given by Eq. \eref{Ktrans} to both the primed and unprimed coordinates in Eq. \eref{Green0}.  Since $\textbf{x}_{\bkappa}=\textbf{x}_0$, one may write $\textbf{R}=\textbf{R}_0=\textbf{R}_{\bkappa}$.  The resulting retarded Green's function is given by:
\begin{equation}
\label{GreenK}
 G(\textbf{x}'_{\bkappa},t'_{\bkappa},\textbf{x}_{\bkappa},t_{\bkappa})=\frac{\delta\left (t'_{\bkappa}-t_{\bkappa}+\displaystyle{\frac{R}{c}}\left (1-\bkappa\cdot\hat{\textbf{R}}\right )\right )}{4\pi R}.
 \end{equation}
This function satisfies the following wave equation:
\begin{equation}
 \left (\frac{1}{c^2}\frac{\partial^2}{\partial t_{\bkappa}^2}-\nabla_{\bkappa}^2-\mathcal{P}_{\bkappa}\frac{\partial}{\partial t_{\bkappa}}\right ) G(\textbf{x}'_{\bkappa},t'_{\bkappa},\textbf{x}_{\bkappa},t_{\bkappa})=\delta\left (t'_{\bkappa}-t_{\bkappa}\right )\delta\left (\textbf{x}'_{\bkappa}-\textbf{x}_{\kappa}\right ).
\end{equation}
In order to obtain the retarded particular solution of Eq. \eref{WaveEqnAni}, one may multiply the expression above by $\psi(\textbf{x}'_{\bkappa},t'_{\bkappa})$ and integrate the primed coordinates over time and space.  It follows that the retarded particular solution is given by:
\begin{equation}
\label{PsiSol}
\Psi(\textbf{x}_{\bkappa},t_{\bkappa})=\frac{1}{4\pi} \int_{R^3}d^3x'_{\bkappa}\frac{\left [\psi\right ]_{\bkappa}}{R}.
\end{equation}
Here I have introduced the following notation (similar to that used by Heras \cite{Hera2007} and Jackson \cite{Jack1999b}):
\begin{equation}
\left [\psi\right ]_{\bkappa}\left (\textbf{x}'_{\bkappa}, \textbf{x}_{\bkappa},t_{\bkappa}\right )
=
\int_{-\infty}^{\infty} dt'_{\bkappa} \delta\left(t'_{\bkappa}-t_{\bkappa}+\frac{R}{c}\left (1-\bkappa\cdot\hat{R}\right ) \right )\psi(\textbf{x}'_{\bkappa},t'_{\bkappa}).
\end{equation}
Using this square bracket notation, the function $\psi$ is evaluated at the retarded time.  As such, the interaction is mediated by signals that travel with the anisotropic velocity given by Eq. \eref{COneway}.

We now have an anisotropic wave equation and its retarded particular solution.  In the next section I will apply these results to the case where the function $\psi$ represents electric charge or current density in order to obtain Maxwell's equations for an observer in absolute motion.

\section{Maxwell's Equations}
\label{MaxwellEqs}
Let's consider the charge continuity equation and determine its behaviour under simultaneity gauge transformation. Using Eq. \eref{Koperators}, one may write the charge continuity equation for $\bkappa\ne 0$ as follows:
\begin{equation}
\label{Continuity}
0=\frac{\partial \rho_0}{\partial t_0}+\nabla_0\cdot \textbf{J}_0 
= \frac{\partial}{\partial t_{\bkappa}}\left (\rho_{0}-\frac{\bkappa\cdot \textbf{J}_0}{c}\right )+\nabla_{\bkappa}\cdot \textbf{J}_0.
\end{equation}
The transformation of $\rho_{0}$ and $\textbf{J}_{0}$ resembles the space-time transformations given by Eq. \eref{Ktrans}.  This affirms the four-vector relationship between charge and current density. 

Eq. \eref{Continuity} immediately suggests a difficulty that is briefly discussed in both \cite{Ande1998} and \cite{Gian1978}.  Suppose, using standard simultaneity, that we have a non-zero electric current $\textbf{J}_0$ and a null electric charge $\rho_0$.  It may seem strange that a non-zero charge density $\rho_{\bkappa}=-\bkappa\cdot\textbf{J}_0/c $ appears when we adjust the simultaneity gauge.  However, if current $\textbf{J}_0$ is present without an accompanying charge density, it follows that $\rho_0$ and $\textbf{J}_0$ are in fact the sum of at least two species of charge and current densities.  These contributions cancel out to give $\rho_0=0$.  However, ``the remote timing operations which must be used to gate the (moving) charge in a given length are affected [by simultaneity gauge], and in an anisotropic manner"\cite{Ande1998}.  As such the cancellation of charge density is synchrony-dependent and will not hold for all values of $\bkappa$. 

We are now equipped with both an anisotropic wave equation Eq. \eref{WaveEqnAni} and a charge continuity equation Eq. \eref{Continuity} for non-standard simultaneity gauges.  The next task is to obtain Maxwell's equations for $\bkappa\ne 0$ using these two elements.  To begin, let's identify $\psi$ in Eq. \eref{WaveEqnAni} with electric charge density $\rho_{\bkappa}$ and then with current density $\textbf{J}_{\bkappa}$:
\begin{eqnarray}
\label{PhiEqn} 
\left (\frac{1}{c^2}\frac{\partial^2}{\partial t_{\bkappa}^2}-\nabla_{\bkappa}^2\right )\phi_{\bkappa}&=\alpha \rho_{\bkappa}+\mathcal{P}_{\bkappa}\frac{\partial \phi_{\bkappa}}{\partial t_{\bkappa}};
\\ 
\label{AEqn}
\left (\frac{1}{c^2}\frac{\partial^2}{\partial t_{\bkappa}^2}-\nabla_{\bkappa}^2\right )\textbf{A}_{\bkappa}&=\beta \textbf{J}_{\bkappa}+\mathcal{P}_{\bkappa}\frac{\partial \textbf{A}_{\bkappa}}{\partial t_{\bkappa}}.
\end{eqnarray} 
According to Eq. \eref{PsiSol}, the retarded particular solutions are given by:
\begin{eqnarray}
\label{PhiSol}
\phi_{\bkappa}(\textbf{x}_{\bkappa},t_{\bkappa})&=\frac{\alpha}{4\pi} \int_{R^3}d^3x'_{\bkappa}\frac{\left [\rho_{\bkappa}\right ]_{\bkappa}}{R};
\\
\label{ASol}
\textbf{A}_{\bkappa}(\textbf{x}_{\bkappa},t_{\bkappa})&=\frac{\beta}{4\pi} \int_{R^3}d^3x'_{\bkappa}\frac{\left [\textbf{J}_{\bkappa}\right ]_{\bkappa}}{R}.
\end{eqnarray}
The constants $\alpha$ and $\beta$ have been introduced as arbitrary parameters to be fixed by choice of units and experiment.  Both wave equations are independent of $\alpha$ and $\beta$ upon division.

One may be tempted to identify $\phi_{\bkappa}$ and $\textbf{A}_{\bkappa}$ with the physical electromagnetic Lorenz potentials.  This would be premature, however, since they are purely mathematical fields representing special cases of $\Psi$ corresponding to $\psi=\alpha\rho_{\bkappa}$ or $\psi=\beta\textbf{J}_{\bkappa}$.  For example, $\rho_{\bkappa}$ could instead represent a number density of non-interacting dust particles and $\textbf{J}_{\bkappa}$ the dust current \cite{Miha1999}.  In such a case there would be no force between the particles despite the existence of mathematical fields $\phi_{\bkappa}$ and $\textbf{A}_{\bkappa}$.  The physical fields are only obtained if the model can be made to fit the experimental data through an appropriate choice of the values of $\alpha$, $\beta$, and $c$. 

Nonetheless, it is fairly straightforward to show that the mathematical fields $\phi_{\bkappa}$ and $\textbf{A}_{\bkappa}$ satisfy a Lorenz condition.  One may use Eqs. \eref{PhiEqn} and \eref{AEqn} to replace $\rho_{\bkappa}$ and $\textbf{J}_{\bkappa}$ in the continuity equation Eq. \eref{Continuity} as follows: 
\begin{equation}
\label{LorenzStep1}
\left (\frac{1}{c^2}\frac{\partial^2}{\partial t_{\bkappa}^2}-\nabla_{\bkappa}^2-\mathcal{P}_{\bkappa}\frac{\partial}{\partial t_{\bkappa}}\right )\left (\frac{\partial \phi_{\bkappa}}{\partial t_{\bkappa}}+\frac{\alpha}{\beta}\nabla_{\bkappa}\cdot \textbf{A}_{\bkappa}\right )=0.
\end{equation}
Here we have a wave operator acting on a function that is either null or a homogeneous solution of the operator.  Since both  $\phi_{\bkappa}$ and $\textbf{A}_{\bkappa}$ are particular solutions of the wave equation it follows that the function must be null.  This is the Lorenz condition: 
\begin{equation}
\label{LorenzCond} 
0=\frac{\partial \phi_{\bkappa}}{\partial t_{\bkappa}}+\frac{\alpha}{\beta}\nabla_{\bkappa}\cdot \textbf{A}_{\bkappa}.
\end{equation}
One may also verify that Eq. \eref{LorenzCond} is true by direct evaluation using Eqs. \eref{PhiSol}, \eref{ASol}, and the continuity equation Eq. \eref{Continuity}.

This Lorenz condition permits one to express the right hand sides of Eqs. \eref{PhiEqn} and \eref{AEqn} in terms of effective electric charge and current density: 
\begin{equation}
\label{EffectiveRhoandJ} \rho_{eff}=\rho_{\bkappa}-\nabla_{\bkappa}\cdot \textbf{P}_{\bkappa}; \qquad
\textbf{J}_{eff}= \textbf{J}_{\bkappa}+\frac{\partial \textbf{P}_{\bkappa}}{\partial t_{\bkappa}}.
\end{equation}
The electric polarization vector $\textbf{P}_{\bkappa}$ is related to the Lorenz vector potential as follows:
\begin{equation}
\label{PConst} \textbf{P}_{\bkappa}=\frac{\mathcal{P}_{\bkappa}\textbf{A}_{\bkappa}}{\beta}.
\end{equation}
Therefore, if Bob shares absolute simultaneity relations with Alice via a non-zero choice of $\bkappa$, he will find the medium through which he travels to be electrically polarized according to Eq. \eref{PConst}.  He must account for both polarization charge and current densities in his electrodynamics using Eq. \eref{EffectiveRhoandJ}. Alice, on the other hand, considers herself to be at absolute rest and sets $\bkappa=0$.  For Alice, the medium is isotropic and $\textbf{P}_{0}=0$.  

One may easily obtain Maxwell's equations as follows.  Eq. \eref{PhiEqn} is equivalent to Gauss' law.  Simply apply the Lorenz condition Eq. \eref{LorenzCond} and rearrange:
\begin{eqnarray}
\alpha \rho_{eff} &=\frac{1}{c^2}\frac{\partial}{\partial t_{\bkappa}}\left (\frac{\partial \phi_{\bkappa}}{\partial t_{\bkappa}}\right )-\nabla_{\bkappa}^2\phi_{\bkappa}
\nonumber \\
&=\frac{1}{c^2}\frac{\partial}{\partial t_{\bkappa}}\left (-\frac{\alpha}{\beta}\nabla_{\bkappa}\cdot \textbf{A}_{\bkappa}\right )-\nabla_{\bkappa}^2\phi_{\bkappa}
\nonumber \\
&=\nabla_{\bkappa}\cdot \left( -\nabla_{\bkappa}\phi_{\bkappa}-\frac{\alpha}{\beta c^2}\frac{\partial \textbf{A}_{\bkappa}}{\partial t_{\bkappa}}\right ). \label{Gauss}
\end{eqnarray}
Similarly, Eq. (16) is equivalent to Ampere's law.  One may use the identity $\nabla_{\bkappa}^2 \textbf{A}_{\bkappa}= \nabla_{\bkappa}\left (\nabla_{\bkappa}\cdot \textbf{A}_{\bkappa}\right )-\nabla_{\bkappa}\times \left (\nabla_{\bkappa}\times \textbf{A}_{\bkappa}\right )$, apply the Lorenz condition Eq. (20), and rearrange:
\begin{eqnarray}
\beta \textbf{J}_{eff} &=\frac{1}{c^2}\frac{\partial}{\partial t_{\bkappa}}\left (\frac{\partial \textbf{A}_{\bkappa}}{\partial t_{\bkappa}}\right )-\nabla_{\bkappa}^2\textbf{A}_{\bkappa}
\nonumber \\
&= \frac{1}{c^2}\frac{\partial}{\partial t_{\bkappa}}\left (\frac{\partial \textbf{A}_{\bkappa}}{\partial t_{\bkappa}}\right )
-\nabla_{\bkappa}\left (\nabla_{\bkappa}\cdot \textbf{A}_{\bkappa}\right )+\nabla_{\bkappa}\times \left (\nabla_{\bkappa}\times \textbf{A}_{\bkappa}\right )
\nonumber \\
&= \frac{1}{c^2}\frac{\partial}{\partial t_{\bkappa}}\left (\frac{\partial \textbf{A}_{\bkappa}}{\partial t_{\bkappa}}\right )
-\nabla_{\bkappa}\left (-\frac{\beta}{\alpha}\frac{\partial\phi_{\bkappa}}{\partial t_{\bkappa}}\right )+\nabla_{\bkappa}\times \left (\nabla_{\bkappa}\times \textbf{A}_{\bkappa}\right )
\nonumber \\
&= \nabla_{\bkappa}\times \left (\nabla_{\bkappa}\times \textbf{A}_{\bkappa}\right )
-\frac{\beta}{\alpha}\frac{\partial}{\partial t_{\bkappa}} \left( -\nabla_{\bkappa}\phi_{\bkappa}-\frac{\alpha}{\beta c^2}\frac{\partial \textbf{A}_{\bkappa}}{\partial t_{\bkappa}}\right )
 \label{Ampere}
\end{eqnarray}
These equations suggest the following standard definitions for electric and magnetic fields:
\begin{equation}
\label{EBDef}
\textbf{E}_{\bkappa}= -\nabla_{\bkappa}\phi_{\bkappa}-\frac{\alpha}{\beta c^2}\frac{\partial \textbf{A}_{\bkappa}}{\partial t_{\bkappa}};
\qquad
\textbf{B}_{\bkappa}=\nabla_{\bkappa}\times \textbf{A}_{\bkappa}.
\end{equation}
The remaining two of Maxwell's equations are automatically satisfied.  Therefore, from Bob's perspective in absolute motion with $\bkappa\ne 0$, Maxwell's equations are given by:
\begin{eqnarray}
\label{MaxGauss} \nabla_{\bkappa}\cdot \textbf{E}_{\bkappa}=\alpha \left(\rho_{\bkappa}-\nabla_{\bkappa}\cdot \textbf{P}_{\bkappa}\right ); 
\\
\label{MaxMonoP} \nabla_{\bkappa}\cdot \textbf{B}_{\bkappa}=0; 
\\
\label{MaxFaraday} \nabla_{\bkappa}\times \textbf{E}_{\bkappa}+\frac{\alpha}{\beta c^2}\frac{\partial \textbf{B}_{\bkappa}}{\partial t_{\bkappa}}=0;
\\ 
\label{MaxAmpere} \nabla_{\bkappa}\times \textbf{B}_{\bkappa}-\frac{\beta}{\alpha}\frac{\partial \textbf{E}_{\bkappa}}{\partial t_{\bkappa}}
=\beta \left (\textbf{J}_{\bkappa}+\frac{\partial \textbf{P}_{\bkappa}}{\partial t_{\bkappa}}\right ).
\end{eqnarray}
These equations do not suffer from the ``unpleasant asymmetric form" of those derived by Rizzi \etal \cite{Rizz2004} since the anisotropy is described using electric polarization $\textbf{P}_{\bkappa}$ (or effective charges and currents $\rho_{eff}$ and $\textbf{J}_{eff}$).

To summarize, one may derive equations resembling Maxwell's equations on the basis of the continuity of charge and the anisotropic wave equation.  The next task is to determine the values of the constants $\alpha$, $\beta$, and $c$ in order to establish that $\textbf{E}_{\bkappa}$ and $\textbf{B}_{\bkappa}$ indeed represent electromagnetic fields.

\section{Dynamics}
\label{Dynamics}
In this section I will discuss how one might fit the model presented above to the experimental data.  Electromagnetism may be understood as a force between conserved sources, namely the transfer of momentum from one source to another through the fields.  Momentum must be conserved within the field system in the absence of the sources.  This is the difference between physical and merely mathematical fields.  Conserved electric charges come with electromagnetic fields that have actual energy and momentum.  Non-interacting yet conserved dust particles come with mathematical electromagnetic fields but these fields will not have any energy and there will be no interaction between the dust particles.

In order to determine values of the constants $\alpha$, $\beta$, and $c$ in the mathematical model developed above, let's consider the following conservation law directly implied by Maxwell's equations Eqs. \eref{MaxGauss}-\eref{MaxAmpere} (for the derivation please see \cite{Grif1999}):
\begin{eqnarray}
\fl-\left (\rho_{eff}\frac{c^2\textbf{E}_{\bkappa}}{\alpha}+\textbf{J}_{eff}\times \frac{\textbf{B}_{\bkappa}}{\beta}\right ) 
&= \frac{\partial}{\partial t_{\bkappa}} \left (\frac{\textbf{E}_{\bkappa}\times \textbf{B}_{\bkappa}}{\alpha\beta}\right ) 
\nonumber \\ \label{LorentzForce}
&+ \nabla_{\bkappa}\cdot \left \lbrace
\frac{I}{2}\left (\frac{c^2\textbf{E}_{\bkappa}^2}{\alpha^2}+\frac{\textbf{B}_{\bkappa}^2}{\beta^2}\right )
-\frac{c^2\textbf{E}_{\bkappa}\textbf{E}_{\bkappa}}{\alpha^2} 
-\frac{\textbf{B}_{\bkappa}\textbf{B}_{\bkappa}}{\beta^2} 
\right \rbrace.
\end{eqnarray}
This conservation law is entirely independent of the constants $\alpha$ and $\beta$ since $\textbf{E}_{\bkappa}$ and $\textbf{B}_{\bkappa}$ are by definition proportional to $\alpha$ and $\beta$ respectively.  The vector field $\textbf{E}_{\bkappa}\times\textbf{B}_{\bkappa}$ is locally conserved in the absence of effective charges $\rho_{\bkappa}-\nabla_{\bkappa}\cdot \textbf{P}_{\bkappa}$ and effective currents $\textbf{J}_{\bkappa}+\partial \textbf{P}_{\bkappa}/\partial t_{\bkappa}$.  One may therefore suppose that field momentum density is proportional to the vector field $\textbf{E}_{\bkappa}\times\textbf{B}_{\bkappa}$.  As a result, the left hand side of the equation above is proportional to a Lorentz force density.

Next, let's consider the constants $\alpha$, $\beta$,  and $c$ by way of a brief diversion. In both \cite{Hera2010ajp} and \cite{Hera2010ejp}, Heras proposes a `$c$ equivalence principle', namely ``the speed $c_u$ obtained in the process of defining electromagnetic units via [instantaneous] action-at-a-distance forces is equivalent to the propagation speed $c$ of electromagnetic waves in vacuum"\cite{Hera2010ajp}.
Heras suggests that it is not necessarily true that $c^2=\alpha/\beta$; rather it just happens to be actually true in nature, much like the numerical equivalence of inertial and gravitational mass.  As such, Heras supposes that ``testing the $c$ equivalence principle to confirm its validity with very high precision would be an interesting and important task"\cite{Hera2010ejp}.  I think that the form of Eq. \eref{LorentzForce} indicates otherwise. The $c$ equivalence principle follows necessarily on this model to within a choice of units.

To show why, notice that the wave equation and continuity of charge are both independent of $\alpha$ and $\beta$. As a result, $\alpha$ and $\beta$ can be divided out of the Lorentz force density on the left hand side of Eq. \eref{LorentzForce}.  Not so with $c$.  This constant determines both the speed of the wave solutions of Eq. \eref{WaveEqnAni} and the ratio of the rates at which effective charges and effective currents obtain momentum from the fields in Eq. \eref{LorentzForce}.  Indeed, one may assume that $c$ governs wave-speed and show that this implies that it also governs the Lorentz force law. Therefore Heras' $c$ equivalence principle follows necessarily from the wave equation and charge continuity.

In any case, let's suppose with Heras that it is (in principle) possible that $c^2\ne\alpha/\beta$.  One is certainly free to introduce two new constants $\alpha'$ and $\beta'$ such that $c^2=\alpha'/\beta'$.  By trivial redefinition, without loss of generality, one may set $\textbf{E}_{\bkappa}'=\alpha'\left (\textbf{E}_{\bkappa}/\alpha\right )$ and $\textbf{B}_{\bkappa}'=\beta'\left (\textbf{B}_{\bkappa}/\beta\right )$ in Eq. \eref{LorentzForce}.  To within a choice of units, it is $\textbf{E}_{\bkappa}'$ and $\textbf{B}_{\bkappa}'$ that represent the electric and magnetic fields, rather than $\textbf{E}_{\bkappa}$ and $\textbf{B}_{\bkappa}$.  Dropping the primes, one may say that if $\textbf{E}_{\bkappa}$ and $\textbf{B}_{\bkappa}$ represent a force per unit charge and current respectively, then it is necessary that $c^2=\alpha/\beta$ (for further discussion of units please see \cite{Jack1999c}).  
 
Returning to the main argument of the paper, it is clear that one specifies either $\alpha$, $\beta$, or $c$ by fixing the units of the system.  A second constant is determined by $c^2=\alpha/\beta$. The third is determined experimentally.  In order to measure the last constant one may conduct either an optical or dynamical experiment.  If the experiment requires measurements at multiple locations then one must specify a value of $\bkappa$ in order to establish a framework of simultaneity in the laboratory.  For optical experiments, the value of $\bkappa$ will determine the electromagnetic wave speed according to Eq. \eref{COneway}.  In the case of a dynamical experiment involving the force between charges or currents, the value of $\bkappa$ will influence the form of Newton's second law.  Details may be found in \cite{Ande1998} or \cite{Sone2009} (for a recent controversy on this issue, please refer to \cite{Ohan2004}).  Although $\bkappa=0$ is likely the simplest---and certainly the most familiar---framework within which to test for the remaining constant, one is free to choose a non-zero value of $\bkappa$ just as one is free to choose a different gauge for the electromagnetic potentials \cite{Ande1998,Ande1992,Ming2003}.  

\section{Discussion}
We have seen that simultaneity may be understood as a gauge, subject to adjustment without experimental consequences.  By choosing to set $\bkappa=-\textbf{v}/c$, Bob can share absolute simultaneity relations with Alice (provide that there is some warrant to regard Alice as a preferred observer \cite{Crai2008,Crai2001}).  From Bob's perspective using a $\bkappa\ne 0$ gauge, one may obtain Maxwell's equations and the Lorentz force law given merely charge continuity and the anisotropic wave equation. 

When Bob considers himself to be in absolute motion (and chooses $\bkappa\ne 0$) he must account for the apparent electric polarization of the vacuum.  This result differs from Ref. \cite{Pucc2005} in which the author neglects to include the field $\textbf{P}_{\bkappa}$ when attempting to derive absolute simultaneity transformations of the electromagnetic fields.  The electric polarization described by Eqs. \eref{Pkdef} and \eref{PConst} is non-linear.  It depends on both the Lorenz vector potential and Bob's velocity relative to Alice, the preferred observer.  The electric polarization field ensures that light travels with the anisotropic velocity given by Eq. \eref{COneway}.  

In light of this demonstration, it ought to be clear that there is no contradiction between the concept of absolute simultaneity and classical electrodynamics (see also \cite{Ande1998,Tang2009,Gian1978}).  Einstein's $\bkappa=0$ stipulation does not play an empirically critical role since it merely represents a convenient choice of simultaneity gauge. The question of absolute simultaneity ought to be assessed on its own merit rather than dismissed on the basis of electrodynamics or special relativity.  

Given the compatibility of absolute simultaneity and electrodynamics, the discussion must turn to whether Alice can truly qualify as a preferred observer.  In the introduction I briefly sketched some potential physical candidates for absolute simultaneity, namely those defined by the cosmic microwave background and non-local quantum correlations.  Further discussion of the merits and demerits of these candidates is beyond the scope of this paper.  However, the demonstration above ought to clear away some of the common misconceptions that tend to obscure such discussion.  Lastly, the philosophical case for absolute simultaneity (see for example \cite{Crai2008,Crai2001}) ought not to be inhibited by concerns about electrodynamics.  By regarding the isotropy of the speed of light as gauge dependent rather than a universal principle, it is much easier to conceive of a world governed by relations of absolute simultaneity.

\ack
The author would like to thank Napoleon Gauthier, Michael Tymchak, and Steven Adam for their assistance.

\end{document}